\documentclass[aps,pre,twocolumn,groupedaddress]{revtex4-1}
\usepackage{amsmath,amssymb,graphicx,tikz,subfigure}
\usepackage[english]{babel}
\usepackage[utf8]{inputenc}
\usepackage{stmaryrd}

\newcommand{\G}{\Gamma}
\newcommand{\hH}{\hat{H}}
\newcommand{\hU}{\hat{U}}
\newcommand{\pa}{\partial}
\newcommand{\eps}{\epsilon}

\newcommand{\la}{\lambda}
\newcommand{\lam}{\lambda_\mathrm{max}}

\newcommand{\tf}{t_\mathrm{f}}
\newcommand{\om}{\omega}
\newcommand{\Om}{\Omega}

\begin{document}

\title{Work statistics in the periodically driven quartic oscillator: classical versus quantum dynamics}

\author{Mattes Heerwagen}
\author{Andreas Engel}

\affiliation{Universit\"at Oldenburg, Institut f\"ur Physik, 26111 Oldenburg, Germany}

\begin{abstract}
In the thermodynamics of nanoscopic systems the relation between classical and quantum mechanical description is of particular importance. To scrutinize this correspondence we study an anharmonic oscillator driven by a periodic external force with slowly varying amplitude both classically and within the framework of quantum mechanics. The energy change of the oscillator induced by the driving is closely related to the probability distribution of work for the system. With the amplitude $\lambda(t)$ of the drive increasing from zero to a maximum $\lam$ and then going back to zero again initial and final Hamiltonian coincide. The main quantity of interest is then the probability density $P(E_f|E_i)$ for transitions from initial energy $E_i$ to final energy $E_f$. In the classical case non-diagonal transitions with $E_f\neq E_i$ mainly arise due to the mechanism of separatrix crossing. We show that approximate analytical results within the pendulum approximation are in accordance with numerical simulations. In the quantum case numerically exact results are complemented with analytical arguments employing Floquet theory. For both classical and quantum case we provide an intuitive explanation for the periodic variation of $P(E_f|E_i)$ with the maximal amplitude $\lam$ of the driving.
\end{abstract}

\pacs{}
\maketitle

\section{Introduction}

Thermodynamics of small systems with typical energy turnover of the order of the thermal energy per degree of freedom builds on probability distributions for the main thermodynamic quantities \cite{Sek10,Jar11,Sei12}. As in macroscopic thermodynamics work and heat are of particular importance and their respective distributions play a pivotal role in stochastic thermodynamics. At the same time neither work nor heat is a state variable, both depend  on the whole process along which a particular state is established. By the first law of thermodynamics they are tied to the energy of the system such that knowledge of one of the two is in general sufficient to determine the other. If the system dynamics are described by classical mechanics there is a clear definition of work as integral of the force along the trajectory. The situation is less clear in the quantum case where no analogue of the classical trajectory exists. Different definitions of work in small quantum systems have been proposed, each with its virtues and drawbacks \cite{Tas00,Yuk00,EspMuk06,CamHaeTal11,TalLutHae07}. 

A first step to cope with the subtleties of defining work in a quantum setting is to confine the attention to closed systems. One then considers a system prepared in equilibrium at inverse temperature $\beta$. Shortly before the process of interest starts system and bath are decoupled from each other. Being isolated during the driving the work performed on the system must be equal to its energy difference. The most obvious way to quantify this difference is to  measure energy before and after the process. This so-called two projective measurement definition of work in a non-equilibrium quantum system is simple and operative. On the down side, the measurements are likely to destroy quantum interferences that may be decisive for the non-classical behaviour of the system. 

To clarify which correlations are destroyed by the two projective measurement prescription and which are kept it is instructive to look in detail at the correspondence between classical and quantum work distributions \cite{GarRonWis17a,GarRonWis17b}. This has been done in \cite{JarQuaRah15} for a quartic oscillator with time-dependent stiffness constant, a simple model system characterized by an integrable classical dynamics. 

The aim of the present paper is to extend this analysis to a quartic oscillator driven by a periodic external force modulated by a slowly varying envelope. This system is interesting for various reasons. Firstly, as typical for driven non-linear oscillators its dynamics show coexistence of integrable and chaotic motion. It is therefore much more representative than a harmonic oscillator with the same driving for which an exact analytical solution is available. Secondly, already on the classical level there are specific mechanisms for depositing energy in the system due to separatrix crossing \cite{DieHenHol92,CarEscTen86}. These transitions show similarities with $\pi$-pulses in quantum systems, which in turn can be understood in terms of constructive or destructive interference of two Floquet states responding adiabatically to the driving envelope~\cite{HolJus94}.

The paper is organized as follows. In section~\ref{sec:system} we define the system and fix the notation. Section~\ref{sec:cl} contains the analysis within classical mechanics. Although all relevant quantities may be expressed analytically the explicit determination of the work distribution and the transition probability requires the numerical solution of the equations of motion. Transforming to action-angle variables of the undriven system we compare our numerical findings to results from the so-called pendulum approximation. Section~\ref{sec:qm} is devoted to the quantum case. Results from the numerical solution of the Schr\"odinger equation are augmented with arguments from Floquet theory. In section~\ref{sec:max} we discuss the oscillatory dependence of the transition probabilities on the maximal amplitude of the driving. Finally, section~\ref{sec:conc} contains our conclusions. 


\section{The system}\label{sec:system}

\begin{figure}
\centering
\includegraphics[width=\linewidth]{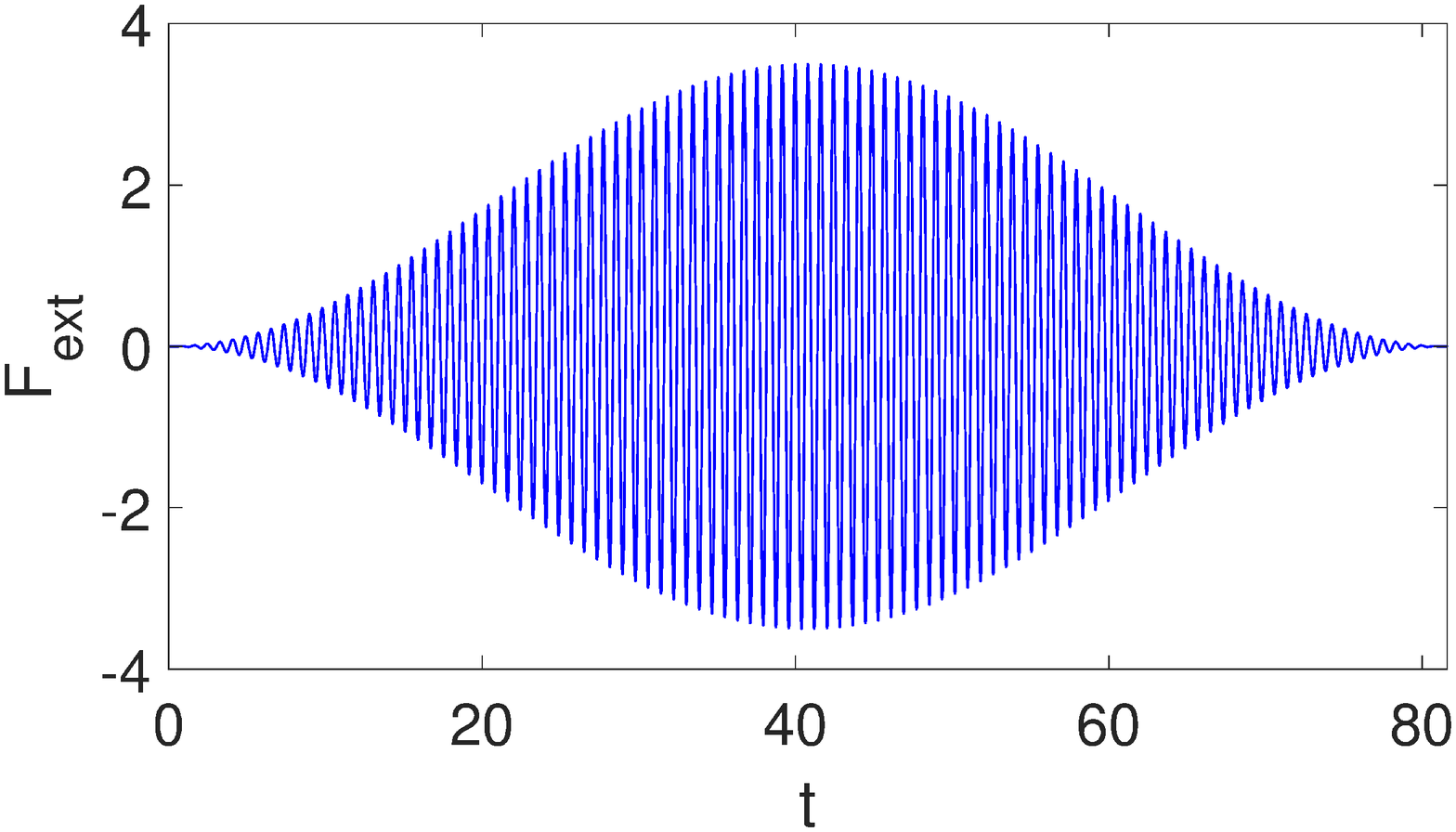}
\caption{External force~\eqref{defF} over time for $\lam=3.5$, $\omega = 7.7$ and $t_f = 100\, 2\pi/\om$.}
\label{extF}
\end{figure}

We consider a particle with mass $m$ moving in one dimension in a potential of the form
\begin{equation}\label{defV}
 V(x)=kx^4
\end{equation} 
at equilibrium with a heat bath at inverse temperature $\beta$. Here $k$ is a parameter that characterizes the strength of the potential. At time $t=0$ we detach the system from the bath and subject it to the time-dependent external force
\begin{equation}\label{defF}
 F_\mathrm{ext}=\la(t)\cos \om t,
\end{equation} 
where $\omega$ is the frequency of the driving, and the envelope function 
\begin{equation}\label{defla}
\la(t) = \lam \sin^2\left(\frac{\pi}{\tf}\, t\right)
\end{equation}
modulates the amplitude of the external force from zero at the beginning of the process through a maximum $\lam$ at $t=\tf/2$, and back to zero at the final time $\tf$. We will always be interested in the case $\tf\gg 2\pi/\om$ with small changes of $\la$ over one period of the driving , see Fig.~\ref{extF}. For our numerical studies we choose
\begin{equation}\label{deftf}
 \tf=1000\, \frac{2\pi}{\om}.
\end{equation} 

Classically, the dynamics is described by the Hamiltonian 
\begin{align}\nonumber
 H(p,x,t) &= \frac{p^2}{2m}+kx^4-x\lambda(t)\cos \om t \\
 &=:H_0-x\lambda(t)\cos \om t, \label{classH} 
\end{align}
where $H_0$ denotes the time-independent part of the Hamiltonian. The quantum analogue of \eqref{classH} is given by 
\begin{equation}\label{quH}
 \hat{H} = \frac{\hat{p}^2}{2m}+k\hat{x}^4-\hat{x}\lambda(t)\cos \om t
         =:\hat{H}_0-\hat{x}\lambda(t)\cos \om t.
\end{equation} 
It fixes the time-evolution operator 
\begin{equation}\label{defU}
\hat{U}(t,0):={\cal T} e^{-\frac{i}{\hbar}\int_0^{t} dt' \hat{H}(t')}
\end{equation}
that describes the unitary dynamics of the system during the driving between $t=0$ and $t=\tf$. Here ${\cal T} $ denotes time-ordering. 

We use 
\begin{equation}\label{dimlessunits}
 \left(\frac{\hbar^2}{2mk}\right)^{1/6}, \quad \frac{(2m)^{2/3}}{(\hbar k)^{1/3}} 
    \quad\mathrm{and}\quad \left(\frac{\hbar^2}{2m}\right)^{2/3} k^{1/3}
\end{equation} 
as units of space, time, and energy, respectively, implying that $p,\la$ and $\om$ are measured in units of 
\begin{equation}
 \hbar^{2/3}(2mk)^{1/6},\quad\hbar\sqrt{\frac{k}{2m}}, 
    \quad\mathrm{and}\quad \frac{(\hbar k)^{1/3}}{(2m)^{2/3}},
\end{equation} 
respectively. The classical equation of motion then reads 
\begin{equation}\label{EOM}
\partial_t^2 x =  2\lambda(t) \cos \om t -8 x^3(t),
\end{equation}
whereas the Schr\"odinger equation determining the time-evolution of the wave function $\psi(x,t)$ acquires the form
\begin{equation}\label{SE}
i\partial_t \psi(x,t)= -\partial_x^2 \psi(x,t) + x^4\psi(x,t) - x \lambda(t) \cos \om t\, \psi(x,t) .
\end{equation}

The interplay between external driving and intrinsic dynamics is most interesting when $\om$ is comparable to the free oscillation frequency of the system. As typical for nonlinear oscillators the oscillation period depends on the amplitude or, equivalently, on the energy. We denote by
\begin{equation}\label{defEom}
 E_\om=\frac{\pi^2}{64\,\big(\Gamma(\frac{3}{4})\big)^8}\, \om^4\simeq 0.03\, \om^4
\end{equation} 
the energy for which the undisturbed particle oscillates with frequency $\om$. To ensure that energies of the order of $E_\om$ are sufficiently likely to occur as initial energies we will mainly choose $\beta=1/E_\om$.

\section{Classical Case} \label{sec:cl}

\subsection{Work distribution}

Classically, the work performed by the external driving is well-defined as integral of the external force along the trajectory of the particle,
\begin{equation}\label{defwcl1}
 W=\int_{x_0}^{x_\mathrm{f}} dx ~ F_\mathrm{ext} =\int_0^{\tf} dt ~ F_\mathrm{ext}\, \pa_t x.
\end{equation} 
Since the force vanishes at the initial and the final time we get after a partial integration
\begin{align}\nonumber
 W&=-\int_0^{\tf} dt ~ x\,\pa_t F_\mathrm{ext}=\int_0^{\tf} dt\, \pa_t H\\\label{defwcl2}
  &=\int_0^{\tf} dt\, \frac{d H}{dt}=H(\tf)-H(0).
\end{align} 
Here we have used the fact that $F_\mathrm{ext}$ is the only time-dependent part of $H$, cf. Eq.~\eqref{classH},  as well as $\pa H/\pa t=dH/dt$ which is well-known from classical mechanics. The work is hence equal to the difference between the final energy $E_f$ and the initial energy $E_i$ in accordance with the first law of thermodynamics and the fact that the system is isolated during the driving. 

Since the system starts at equilibrium the initial energy $E_i$ is a random quantity distributed in accordance with the canonical distribution. We denote this initial distribution by $P^C_i(E)$ where the superscript indicates the classical case. With $E_i$ also the final energy, $E_f$, and the work performed, $W$, will be random quantities. The probability distribution for the work may be written as 
\begin{equation}\label{defPWcl}
 P^C(W) = \int\!\! dE_i \!\!\int \!\! dE_f ~ P^C_i(E_i) P^C_t(E_f|E_i) \delta(W-E_f+E_i),
\end{equation}
where $P^C_t(E_f|E_i)$ denotes the transition probability to end up in energy $E_f$ when started with energy $E_i$. Note that this is a non-trivial quantity even for Hamiltonian systems. Although the mapping from the initial values of $x$ and $p$ to their final ones is deterministic, the determination of $P^C_t(E_f|E_i)$ requires to find the fraction of initial phase space points with $H_0=E_i$ that will end up in a final point with $H_0=E_f$. 

\begin{figure}
\centering
\includegraphics[width=\linewidth]{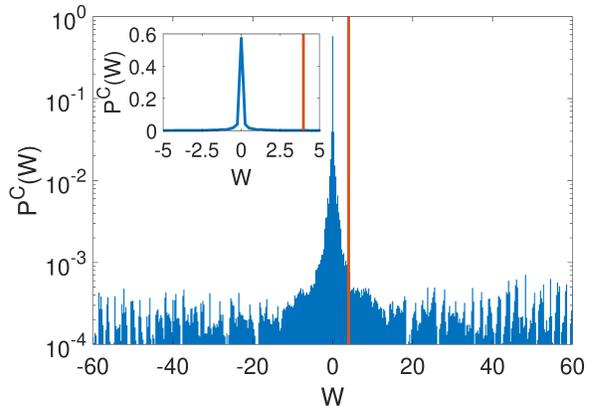}
\caption{Histogram of work values as defined in \eqref{defPWcl} obtained from numerical integrations of the classical equation of motion~\eqref{EOM}. Parameter values are $\lam = 3.5$, $\omega = 7.7$, and $\tf=1000\,2\pi/\om$. The inverse temperature of the bath is $\beta = 1/E_\om$ with $E_\om$ defined by~\eqref{defEom}. The average value resulting from this distribution is indicated by the red line.}
\label{classWork}
\end{figure}

\begin{figure}
\centering
\includegraphics[width=\linewidth]{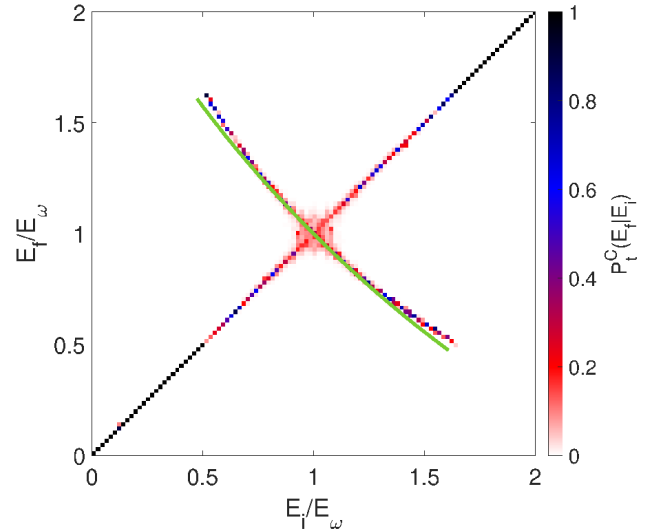}
\caption{Classical transition probability $P^C_t(E_f|E_i)$ color-coded as function of the initial and final energy. The parameters are the same as in Fig.~\ref{classWork}. Outside the transition window $0.5\lesssim E_i/E_\om\lesssim 1.6$ the final energy is almost always identical with the initial one. The green line shows the analytical result from the pendulum approximation.}
\label{classTrans}
\end{figure}

A possible way to determine $P^C_t(E_f|E_i)$ is by sampling the initial energy shell microcanonically. To do so one picks  points on this shell at random with equal probability, uses them as initial condition for a numerical integration of the equation of motion, and determines the final value of the energy. Fig.~\ref{classWork} shows an example of a  work distribution obtained in this way; in Fig.~\ref{classTrans} the corresponding transition probability $P^C_t(E_f|E_i)$ is displayed. 

The work distribution is rather concentrated around $W=0$ corresponding to $E_f=E_i$ as can be clearly seen in the inset. The logarithmic plot shown in the main figure, however, demonstrates that $P^C(W)$ has pronounced tails to rather large values of $|W|$. These tails extend even beyond the interval of $W$ shown. The strong fluctuations and gaps in these tails are due to the finite number of sampling points implemented. 

From the inset one may have the impression that the distribution $P^C(W)$ is symmetric around $W=0$. Nevertheless, the average value $\langle W\rangle$ indicated by the red line in the figure is positive and markedly different from the most probable values of $W$. This is a consequence of the Jarzynski equality \cite{Jar97}
\begin{equation}\label{JE}
 \langle e^{-\beta W}\rangle=e^{-\beta \Delta F}=1
\end{equation} 
where the last equality follows from the fact that in the present case the free-energy difference $\Delta F$ is zero due to $H(\tf)=H(0)=H_0$. Consistently, the histogram of work values shown in Fig.~\ref{classWork} yields

\begin{equation}
\langle e^{-\beta W}\rangle_\mathrm{hist}=1.011.
\end{equation}

By Jensen's inequality Eq.~\eqref{JE} implies $\langle W\rangle\geq 0$. 
From the plot it is also discernible that the differences between the probabilities for positive and negative work values occur mainly in the tails of the distribution. It is well-known that the Jarzynski equality is particularly sensitive to these tails \cite{Jar11}.

The transition probability $P^C_t(E_f|E_i)$ shown in Fig.~\ref{classTrans} has a rather peculiar structure. Outside the interval $0.5\lesssim E_i/E_\om\lesssim 1.6$ hardly any transition to other energy values occur, $P^C_t(E_f|E_i)\sim \delta(E_f-E_i)$. This part of $P^C_t(E_f|E_i)$ therefore contributes almost exclusively to the central peak of $P^C(W)$ at $W=0$. Within this energy window, on the other hand, there is appreciable probability for non-diagonal transitions with $E_f\neq E_i$. Except for a small region surrounding the point $E_i=E_f=E_\om$ these transitions are remarkably concentrated around one particular value of $E_f-E_i=W$. Near the crossing of the two main transition lines of the figure, at $E_i\simeq E_f\simeq E_\om$, the transition probability is smeared out over a small region. In the next section we provide approximate analytical arguments to understand these features of $P^C_t(E_f|E_i)$ qualitatively and quantitatively.


\subsection{Action-angle variables}\label{chAA}

\begin{figure*}
\centering
\includegraphics[width=.4\linewidth]{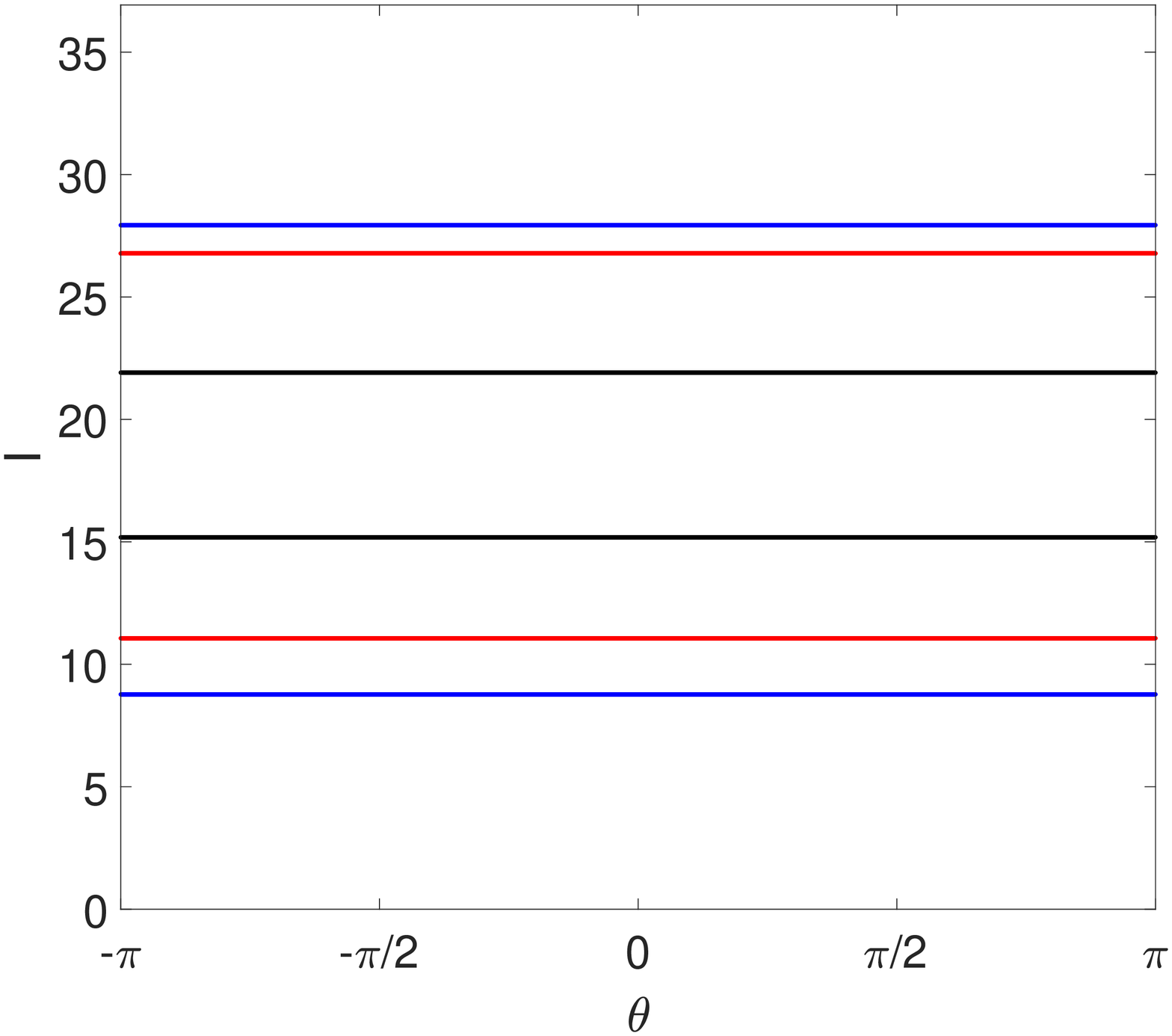}\hspace{.1\linewidth}
\includegraphics[width=.4\linewidth]{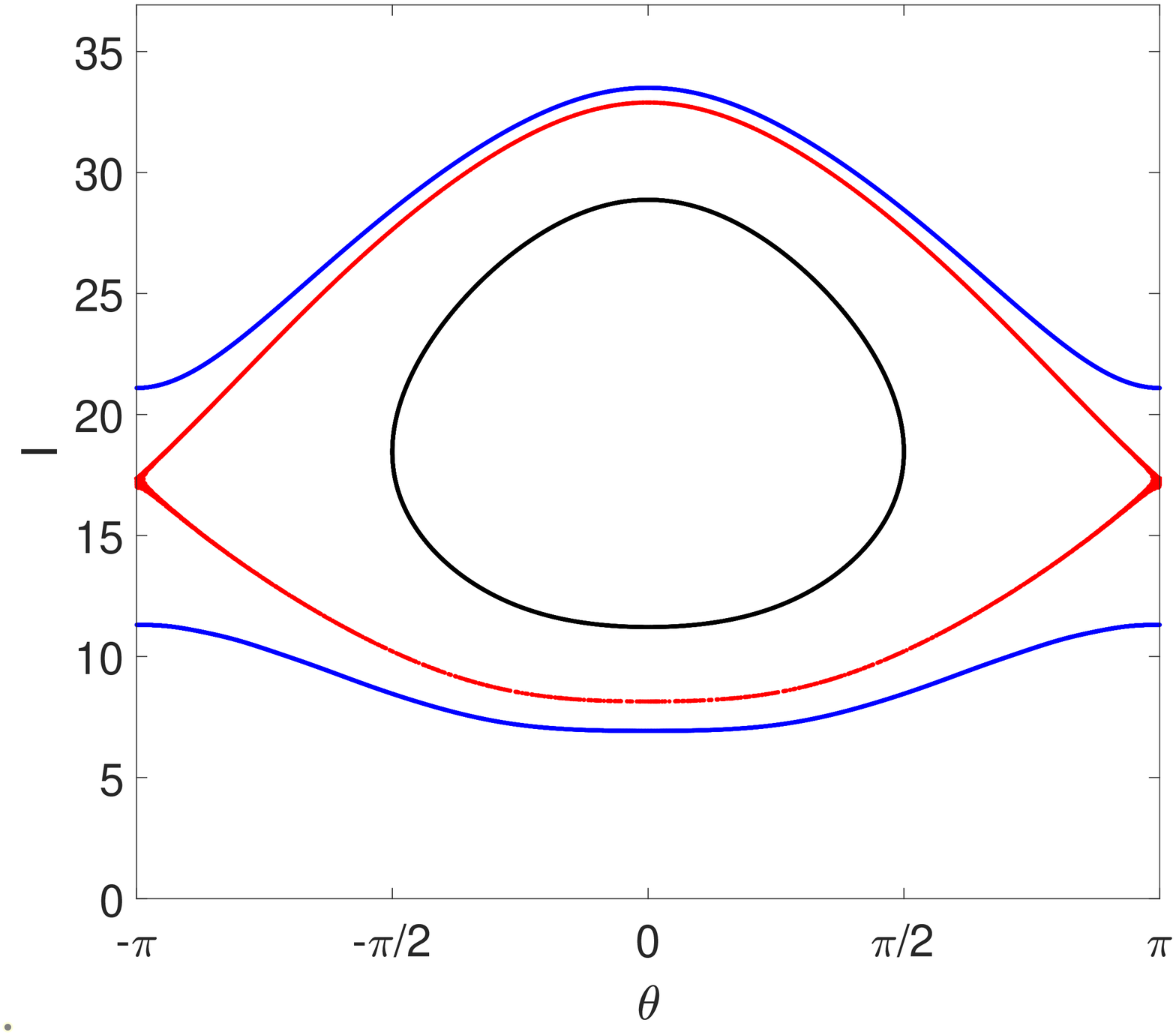}
\caption{Poincar\'e sections showing the dependence of the action-angle variables $(I,\theta)$ of the undriven system under a dynamics with $\la=$const. Left: $\la=0$ implying $I(\theta)=$const. Shown are lines for six  different choices of $E_i$. Right: $\la=\lam$. The six lines $I(\theta)$ are now bend and may meet forming closed loops such that transitions by separatrix crossing to values $E_f\neq E_i$ may take place.}
\label{AA}
\end{figure*}

A transparent qualitative characterization of the classical transition probability $P_t^C (E_f|E_i)$ can be obtained in terms of action-angle variables of the unperturbed system characterized by $H_0$, cf. Eq.~\eqref{classH}. To transform from the initial canonical variables $x$ and $p$ to the 
action-angle variables $I$ and $\theta$ we follow the standard procedure \cite{LanLifI50} and introduce
\begin{equation}\label{defI}
 I := \frac{1}{2\pi} \oint_{H_0} \!\!\!dx~p(x;H_0)=\frac{1}{2\pi} H_0^{3/4} B\left(\frac{1}{4},\frac{3}{2}\right),
\end{equation} 
where the integral is over the classical orbit $p(x)$ with $H_0(x,p)$ staying constant and
\begin{equation}
 B(x,y):=\frac{\G(x)\G(y)}{\G(x+y)}
\end{equation} 
denotes Euler's Beta function \cite{AbrSte72}. Inverting \eqref{defI} we find 
\begin{equation}\label{h1}
H_0(I) = 3^{4/3} \frac{\Gamma\left(\frac{3}{4}\right)^{8/3}}{(2\pi)^{2/3}} I^{4/3}=:C\,I^{4/3},
\end{equation} 
where the various numerical prefactors were subsumed into the constant $C$. For the oscillation frequency $\Om$ of the undriven system this implies
\begin{equation}\label{h2}
 \Om=\frac{\pa H_0}{\pa I}=\frac{4}{3} C\, I^{1/3}.
\end{equation} 

If the frequency of the undriven system, $\Om$, coincides with the frequency of the external drive, $\om$, we are at resonance characterized by the action
\begin{equation}\label{defIom}
 I_\omega := \left(\frac{3\omega}{4C}\right)^3.
\end{equation}

Plugging \eqref{defIom} into \eqref{h1} we find 
\begin{equation}
E_\omega := H_0(I_\omega) = \frac{81}{256\, C^3} \omega^4,
\end{equation}
which is equivalent to \eqref{defEom}.

We may now write the complete Hamiltonian \eqref{classH} in terms of $I$ and $\theta$ to find 
\begin{equation} \label{H_aa}
H(I,\theta,t) = H_0(I) - x(I,\theta) \lambda(t) \cos \om t.
\end{equation}
Here $x(I,\theta)$ is a function fixed by the canonical transformation performed that may be written in terms of Jacobi  elliptic functions. Its explicit form in not needed for what follows. It is only important that this function is periodic and even in $\theta$:
\begin{equation}
 x(I,\theta+2\pi)=x(I,\theta),\qquad x(I,-\theta)=x(I,\theta).
\end{equation} 

It is instructive to consider Poincar\'e sections of the action $I$ as function of the angle $\theta$ for different {\em constant} values of $\la$. To this end we choose initial conditions $(x_i,p_i)$ for the classical equation of motion \eqref{EOM} that correspond to a prescribed value of $I(\theta=0)$ and integrate these equations numerically. From the values of $x$ and $p$ at stroboscopic times $t_n=n\,2\pi/\om$ we determine $I(t_n)$ and $\theta(t_n)$ that for each value of $n$ give rise to one point in the Poincar\'e plots. Fig.~\ref{AA} compares plots generated in this way for $\la=0$ (left) and $\la=\lam$ (right). As indicated by the colors each curve in the right figure derives from a corresponding one of the left figure under slow variations of $\la$ from $\la=0$ to $\la=\lam$.

For $\la=0$ the system is autonomous and integrable and correspondingly the action $I$ is a constant of motion independent of $\theta$. The six lines shown in the left part of Fig.~\ref{AA} correspond to six different values of $I$ and therefore, via \eqref{h1}, to six different values of the system energy $E_i$. As can be seen from the right part of Fig.~\ref{AA}, for $\la=\lam$ the curves bend up or down near $\theta=\pm\pi$ and three qualitatively different types of trajectories can be distinguished. 

The first type is exemplified by the two blue lines. Despite their distortion they remain separated from each other for all values of $\la$ between zero and $\lam$, i.e., when starting at say the lower blue line of the left Figure at $t=0$ and increasing $\la$ slowly no transition to the upper one is likely to occur during the whole driving. Eventually, when $\la=0$ again at $t=\tf$ the systems returns back to the initial value of $I$ and hence also to its initial energy, $E_f=E_i$. Lines of this type, therefore, generate the black diagonal points outside the transition window in Fig.~\ref{classTrans}.

The second type of lines is given by the red ones touching at the boundary $\theta=\pm \pi$ of the $\theta$ domain for just the maximal value $\lam$ of $\la$. They give the stroboscopic picture of the separatrix at $\la=\lam$ since they separate trajectories of the blue type from those of the black one.  For the complete time-dependent process characterized by $\la(t)$ as given by \eqref{defla} this means, that for a short moment at maximal $\la$ transitions between the two red lines may take place. A small fraction of systems starting at $t=0$ on the lower red line may end up at $t=\tf$ on the upper one with different energy, $E_f\neq E_i$. The values of $E_i$ corresponding to the two red lines in Fig.~\ref{AA} therefore define the beginning and the end of the transition window in Fig.~\ref{classTrans}.

Finally, the third type of lines represented by the black pair in Fig.~\ref{AA} stand for energy values inside this transition window. For them there is a value $\la_c$ with \mbox{$0<\la_c<\lam$} such that their shape is similar to the blue lines in Fig.~\ref{AA} for $\la(t)<\la_c$ and like the red ones for $\la(t)=\la_c$. At this point they cross the separatrix and for $\la(t)>\la_c$ trajectories originating from different initial values of $I$ mix on the same closed black line. When, after reaching $\lam$, $\la$ decreases again the trajectories cross the separatrix again and reappear as separated black lines that eventually deform back to their original shape when $\la=0$. Depending on the details of the dynamics several trajectories that started out at the lower value of $I$ will end up in the higher one and vice versa. This mechanism has been dubbed {\em separatrix crossing} in~\cite{CarEscTen86,DieHenHol92} and gives rise to the non-diagonal transitions seen in Fig.~\ref{classTrans}.

Note that the described transitions take place only for two matching values of $I$ and therefore also only for matching pairs of $E_i$ and $E_f$. This explains the peculiar structure of $P^C_t(E_f|E_i)$ shown in Fig.~\ref{classTrans}. Near the resonance region, $E_i\simeq E_\om$, the simple picture of describing the full dynamics with a time-dependent $\la(t)$ in terms of successive Poincar\'e plots corresponding to constant values of $\la$ breaks down and the structure of the transition probability $P^C_t(E_f|E_i)$ becomes richer.


\subsection{Pendulum approximation}

In addition to the qualitative understanding of the transitions obtained in the previous subsection it is possible also to derive an approximate expression for the relation between $E_i$ and $E_f$ in these transitions by invoking the so-called pendulum approximation \cite{Chi79,Hol95}. 

Since $x(I,\theta)$ is a periodic and even function of $\theta$ it may be represented by a Fourier series involving cosine functions only:
\begin{equation}
x(I,\theta) = \sum_{n=0}^\infty x_n(I)\cos n\theta. 
\end{equation}
Here the $x_n(I)$ are given as usual by 
\begin{equation}
x_n(I) = \frac{1}{\pi} \int_{-\pi}^\pi d\theta ~ x(I,\theta)\cos n\theta.
\end{equation}
For {\em constant} $\la$ the Hamiltonian \eqref{H_aa} then acquires the form 
\begin{align}\nonumber
H&(I,\theta,t) = H_0(I) - \lambda\sum_{n=0}^\infty x_n(I) \cos n\theta\,\cos \om t\\\label{h3}
 &= H_0(I) - \frac{\lambda}{2}\sum_{n=0}^\infty x_n(I) \left[\cos(n\theta-\om t)+\cos(n\theta+\om t)\right].
\end{align}

As discussed in the previous subsection transitions to other energy values occur only within an energy window around $E_i=E_\om$, i.e., for values of $I$ not too different from the resonance value $I_\om$. We hence expand $H_0(I)$ up to second order around $I_\om$,  
\begin{align}\label{h4}
H_0(I)= H_0(I_\omega) + \omega (I-I_\omega)+ \frac{1}{2M} (I-I_\omega)^2+...
\end{align}
where we have used \eqref{h2} and introduced the abbreviation 
\begin{equation}\label{defM}
 M := \left( \left.\frac{\partial^2 H_0}{\partial I^2}\right|_{I=I_\omega} \right)^{-1} = \frac{9}{4C}\, I_\omega^{2/3}.
\end{equation} 

Moreover, we only keep the slowly time-dependent resonant term $\cos(\theta - \omega t)$ in \eqref{h3} to obtain the approximate expression  
\begin{align}
\nonumber H(I,\theta,t) \approx H_0(I_\omega) &+ \omega (I-I_\omega) + \frac{1}{2m} (I-I_\omega)^2 \\\label{Hpendapp}
& - \frac{\lambda}{2} x_1(I_\omega) \cos(\theta-\omega t).
\end{align}

Next we perform yet another canonical transformation from $(I,\theta)$ to $(P,\phi)$ defined by the generating function 
\begin{equation}
F(I,\phi,t) = -(I-I_\omega) (\phi+\omega t).
\end{equation}
It gives rise to 
\begin{align}\label{pend1}
P &= -\frac{\pa F}{\pa \phi}=I-I_\omega\\\label{pend2}
\theta&=-\frac{\pa F}{\pa I}=\phi+\om t,
\end{align}
as well as to the new Hamiltonian 
\begin{equation}\label{H_new}
 K(P,\phi)= H+\frac{\pa F}{\pa t}-H_0
         = \frac{1}{2m} P^2 - \frac{\lambda}{2} x_1(I_\omega) \cos\phi,
\end{equation} 
where we have subtracted the irrelevant constant $H_0(I_\omega)$. 

The new Hamiltonian $K$ is not explicitly time-dependent and describes a simple one-dimensional pendulum with mass $M$ and potential 
\begin{equation}\label{defU0}
 U(\phi)=\frac{\lambda}{2} x_1(I_\omega) \cos\phi=: U_0 \cos\phi.
\end{equation} 

The separatrix for the pendulum is given by \mbox{$K=U_0.$} The corresponding value of the momentum is 
\begin{align}\nonumber 
P_\mathrm{sx} &=\pm \sqrt{2M(K+ U_0 \cos\phi)}= \pm \sqrt{2M(U_0 + U_0 \cos\phi)} \\
&= \pm \sqrt{4MU_0}\left| \cos\left(\frac{\phi}{2}\right)\right|.
\end{align}
Transforming back to $I$ and $\theta$ according to \eqref{pend1} and \eqref{pend2} and using the definitions of $U_0$ and $M$ in \eqref{defU0} and \eqref{defM} respectively we find for the corresponding value of $I$
\begin{align}\label{defDI}
I_\mathrm{sx} = I_\omega \pm \Delta I(\lambda) \left|\cos\left(\frac{\theta - \omega t}{2}\right)\right|
\end{align}

with

\begin{equation}
\Delta I(\lambda) = \sqrt{\frac{9 \lambda}{2C} I_\omega^{2/3}\, x_1(I_\omega)}.
\end{equation}

\begin{figure}
\centering
\includegraphics[width=\linewidth]{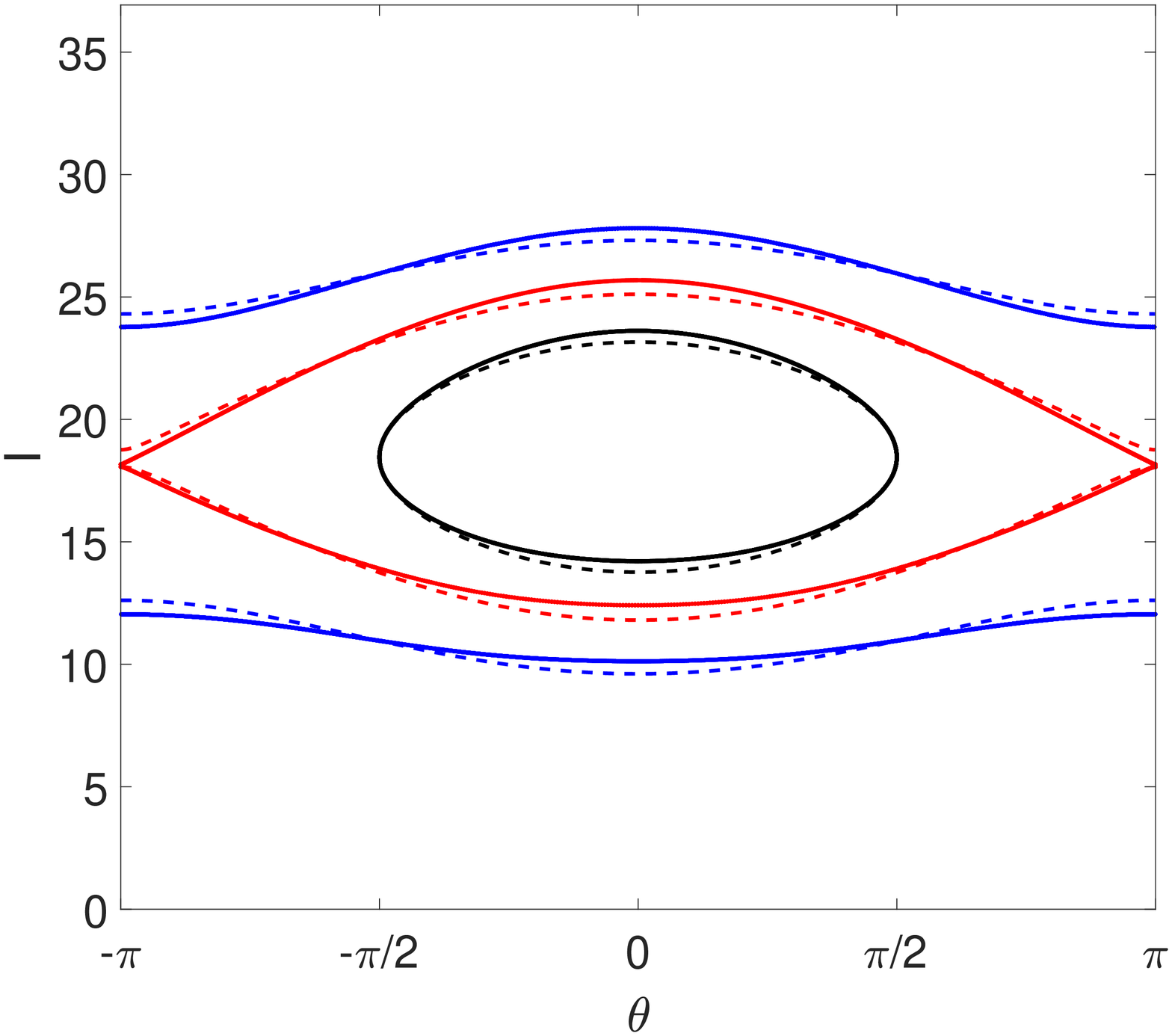}
\caption{Poincar\'e sections for the original system~\eqref{H_aa} (full lines) and for the pendulum approximation~\eqref{Hpendapp} (dashed lines) for $\la=1$. Note that the colors do not correspond to the same values of $E_i$ in Fig.~\ref{AA} since the value of $\la$ is different.}
\label{AA2}
\end{figure}

Fig.~\ref{AA2} compares Poincar\'e plots for the full system~\eqref{H_aa} with those within the pendulum approximation at \mbox{$\la=1$}. The red lines correspond to $I=I_\mathrm{sx}$ as defined in \eqref{defDI}. There is good agreement between the numerical results for the original system and the approximate analytic theory. This agreement is, however, confined to comparatively small values of $\la$. That the pendulum approximation becomes less reliable with increasing $\la$ can already be anticipated by comparing the full lines of Figs.~\ref{AA} right and Fig.~\ref{AA2} corresponding to $\la=\lam=3.5$ and $\la=1$ respectively: The asymmetry between upper and lower parts of the curves increases with $\la$. Since \eqref{defDI} implies a symmetric shape of the corresponding curves for the pendulum approximation the deviation between exact results and analytical approximation necessarily grows with increasing $\la$. This is, of course, also in accordance with the truncated expansion in \eqref{h4}.

Within the pendulum approximation the transitions occur between values $I_i$ and $I_f$ that are located symmetrically around $I_\om$, cf.~\eqref{defDI} . Hence 
\begin{equation}
 I_f=2I_\om -I_i
\end{equation} 
and using \eqref{h1} we find 
\begin{equation}\label{resPpendapprox}
 \frac{E_f}{E_\om}=\left(2-\left(\frac{E_i}{E_\om}\right)^\frac{3}{4}\right)^\frac{4}{3}.
\end{equation} 
This relation is shown as green line in Fig.~\ref{classTrans}. It agrees well with the numerical results from the full dynamics if $E_i$ does not differ too much from $E_\om$, i.e. for small values of $W$. At the border of the transition window in Fig.~\ref{classTrans} larger values of $\la$ dominate the transitions and, as discussed above, the pendulum approximation becomes less accurate.

To determine the size of the transition window within the pendulum approximation, i.e., the points at which the green line in Fig.~\ref{classTrans} starts and ends, we need to find the values of $E_i$ associated with the upper and the lower part of the dotted red line in Fig.~\ref{AA2}. This could be done similarly to Fig.~\ref{AA} by numerically solving the equation of motion corresponding to $K(P,\phi)$ for a slowly decreasing $\la(t)$. It is, however, more direct to use the adiabatic invariance \cite{LanLifI49} of the action $I_K$ of the pendulum Hamiltonian $K(P,\phi)$. To determine the maximal size of the transition window we have to consider the separatrix, i.e. to put $K=U_0$, for the case $\lambda = \lam$:

\begin{align}
\nonumber I_K(\lam) :&=  \frac{1}{2\pi} \oint_K  d\phi~P_\mathrm{sx}(\phi;K) \\
\nonumber &= \pm \frac{1}{2\pi} \sqrt{4MU_0(\lam)} \int_{-\pi}^\pi d\phi \left|\cos \frac{\phi}{2}\right| \\
&= \pm \frac{2}{\pi} \sqrt{4MU_0(\lam)} = \pm\frac{2}{\pi} \Delta I(\lam)
\end{align}

For slow variation of $\la$ these values do not change down to $\la=0$ where they give rise to the two initial values of the action 
\begin{equation}
 I_0 = I_\omega \pm \frac{2}{\pi} \Delta I (\lam) \,. 
\end{equation}
Via \eqref{h1} these two values of $I_0$ determine the boundaries of the transition window within the pendulum approximation.


\section{Quantum case} \label{sec:qm}

\subsection{Work distribution}

As discussed already in the introduction the concept of work for a quantum system is intricate. A definition similar to \eqref{defwcl1} is impossible because there is no quantum analog to the trajectory $x(t)$. In what follows we will use the two projective measurement prescription of work and measure the energy of the system before the driving starts at $t=0$ and a second time immediately after the driving ends at $t=\tf$. The corresponding energy values are again called $E_i$ and $E_f$ respectively and the work is defined as their difference
\begin{equation}
W:=E_f-E_i.
\end{equation}
Although this expression looks deceptively similar to \eqref{defwcl2} two differences must be kept in mind. First, there is no longer a connection with a definition like \eqref{defwcl1}, and second, the difference of the Hamiltonians is replaced by the difference of their measurement values. 

The probability distribution of the work has a form similar to \eqref{defPWcl}
\begin{equation}\label{defPWqu}
P^Q(W) = \sum_{i,f} P_i^Q(E_i) P^Q_t(E_f|E_i)\,\delta_{W,E_f-E_i},
\end{equation}
where the integrals are replaced by sums that run over all initial and final states. 

Similar to the classical case $P_i^Q(E_i)$ is determined by the canonical distribution characterizing the equilibrium state of the system at $t=0$. The first energy measurement projects the state of the system to an energy eigenstate $\left|\phi_i\right>$ of the undriven Hamiltonian $\hH_0$ defined in \eqref{quH} with probability 
\begin{equation}\label{defPiqu}
 P_i^Q(E_i)=\frac{1}{Z}\, e^{-\beta E_i}.
\end{equation}
Here $E_i$ is the eigenvalue corresponding to $\left|\phi_i\right>$ and $Z$ denotes the canonical partition function
 \begin{equation}
 Z:= \sum_n e^{-\beta E_n}.
\end{equation}

The stationary Schr\"odinger equation with quartic potential cannot be solved analytically and we have to determine a characteristic set of eigenvalues and eigenstates numerically. 
For a meaningful comparison with the classical results discussed in section~\ref{sec:cl} eigenstates $\left|\phi_n\right>$ up to $n=40$ are needed. Discretizing the $x$-axis in the interval $-8\leq x\leq 8$ into 4000 points we found the matrix Numerov method \cite{PilGogWal12} an efficient and accurate tool to generate these states together with their eigenvalues.

Contrary to the classical case in which the Hamiltonian dynamics during the driving is deterministic the second energy measurement involves an additional piece of randomness that is of 
genuine quantum nature. It is contained in the transition probability $P^Q_t(E_f|E_i)$ that is again the central quantity of interest.


\subsection{Quantum transition probability}

\begin{figure}
\centering
\includegraphics[width=\linewidth]{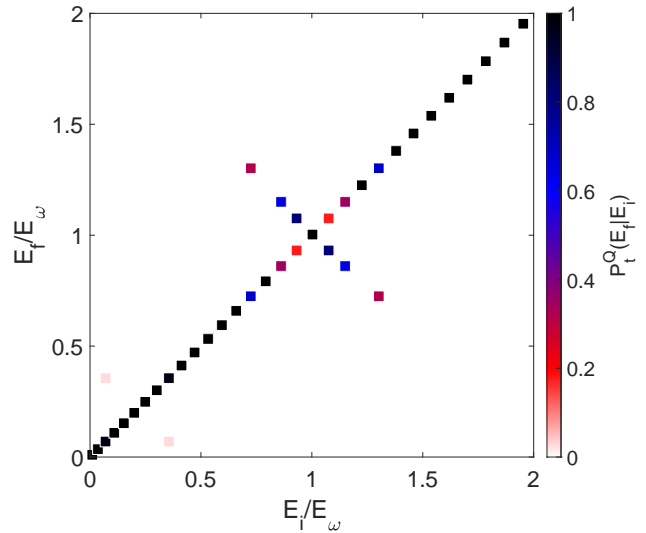}
\caption{Quantum transition probability $P^Q_t(E_f|E_i)$ color-coded as function of the initial and final energy. The parameters are the same as in Fig.~\ref{classTrans}, the time step used in the numerical integration of the Schr\"odinger equation is $\Delta t=4\cdot 10^{-5}$.}
\label{quanTrans}
\end{figure}

To calculate the transition probability $P^Q_t(E_f|E_i)$ we need to know the state 
\begin{equation}
 \left|\phi\right>:= \hU(\tf,0) \left|\phi_i\right>
\end{equation}
to which the system evolves during the driving when started in $\left|\phi_i\right>$ at $t=0$. The second energy measurement then gives rise to 
\begin{equation}
 P^Q_t (E_f|E_i) = \left|\left<\phi_f |\phi\right>\right|^2= \left|\left<\phi_f\right|\hat{U}(\tf,0)\left|\phi_i\right>\right|^2.
\end{equation}
We determine $|\phi\rangle$ from a numerical solution of the time-dependent Schr\"odinger equation using the Crank-Nicolson method \citep{Moy04, Preetal07}. This methods builds on the discretization of the time evolution operator \eqref{defU} in Cayley form
\begin{equation}\nonumber
\hat{U}(t+\Delta t,t)={\cal T} e^{-i\int_t^{t+\Delta t}\!\!dt' \hat{H}(t')}
                = \frac{1-i \hat{H}(t) \frac{\Delta t}{2}}{1+i \hH(t) \frac{\Delta t}{2}} + O(\Delta t^2),
\end{equation}
where $\Delta t$ denotes the temporal step size. The main virtue of this replacement is that the leading term on the r.h.s. is unitary and therefore norm-preserving. 

Fig.~\ref{quanTrans} shows results for the quantum transition probability obtained in this way. The diagonal structure is similar to the classical case shown in Fig.~\ref{classTrans}. Outside an energy window around $E_\om\simeq E_{19}$ there are only few transitions to other energy values and the system mostly returns to its initial energy $E_i$ at $t=\tf$. Inside this window transitions always occur to just one final energy $E_f\neq E_i$ which is very similar to the energy $E_f$ found in the classical case. It is clear that due to the discrete energy spectrum fewer values of $\Delta E$ are realized. There are three important differences between classical and quantum mechanical case. 

First, the quantum transition window is much smaller than the classical one. For the transitions found $\Delta E$ is near to an even multiple of $\om$ and the largest value observed is $\Delta E=8\om$. Due to the fact that the energy spectrum of the quartic oscillator is not equidistant, there is no matching pair of energy eigenstates $i,f$ satisfying $|E_f-E_i|=\Delta E$ for larger even multiples of $\om$. 

Second, the transition with $\Delta E=6\om$, although within the accessible energy window, is missing in the quantum case giving rise to a gap in the secondary diagonal formed by transitions with $E_f\neq E_i$. This can be understood as consequence of destructive interference between Floquet states as we discuss in detail in the next subsection. 

Third, there is a small but non-zero transition probability between the states with $n=3$ and $m=9$ as visible in the lower left corner of Fig.~\ref{quanTrans}. This transition has no classical analogue. Its mechanism can again be understood within the framework of Floquet theory, cf. subsection~\ref{sec:Floquet}.

\begin{figure}
\centering
\includegraphics[width=\linewidth]{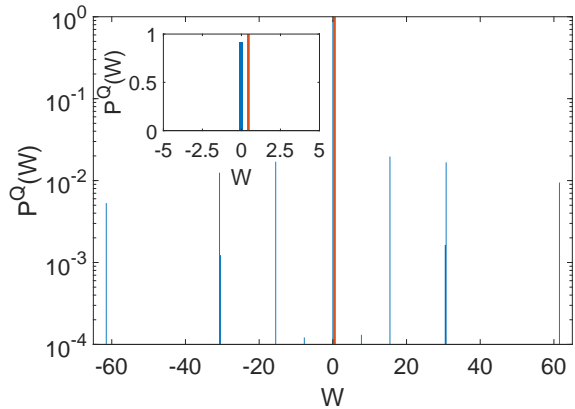}
\caption{Histogramm of work values for the quantum case. The parameters are the same as in Fig.~\ref{classWork}, the time step used in the numerical integration of the Schr\"odinger equation is $\Delta t=4\cdot 10^{-5}$. The average value of the work resulting from the histogram and indicated by the red line is again positive as required by the Jarzynski equality~\eqref{JE}.}
\label{quanWork}
\end{figure}

The quantum work distribution~\eqref{defPWqu} resulting from \eqref{defPiqu} and the numerical determination of $P^Q_t (E_f|E_i)$ is shown in Fig.~\ref{quanWork}. There is  close correspondence with Fig.~\ref{quanTrans}. The central peak at $W=0$ derives from the diagonal transitions in Fig.~\ref{quanTrans}. The six highest peaks at non-zero $W$ correspond to the transitions with $\Delta E=2\om, 4\om$ and $8\om$ on the secondary diagonal in Fig.~\ref{quanTrans}. Again the gap at $W=\pm6\om$ is clearly seen. Finally, the smaller peaks near $W=4\om$ are due to the transitions between states $3$ and $9$. 

Despite looking rather symmetrically around $W=0$ the histogram compiles slightly more probability at positive $W$ so that the average value $\langle W \rangle$ is again larger than zero in accordance with the Jarzynski equality~\eqref{JE}. More precisely, we find from the numerical data

\begin{equation}
\langle e^{-\beta W}\rangle_\mathrm{hist}=0.999.
\end{equation}


\subsection{Floquet Theory}\label{sec:Floquet}

\begin{figure}
\includegraphics[width=\linewidth]{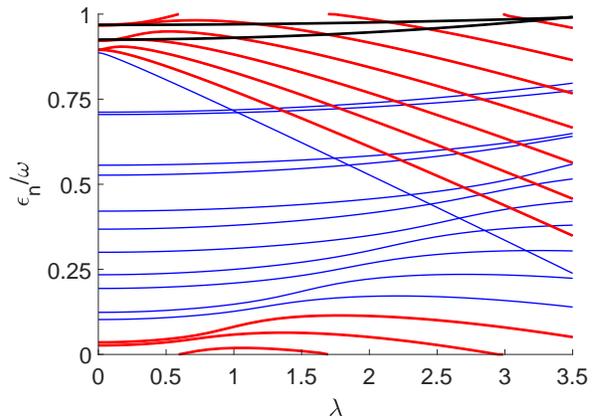}
\caption{Selection of quasienergies $\epsilon_n$ divided by $\om$ for $n=9,...,29$ as function of $\lambda$ for the parameter values of Fig.~\ref{quanTrans}.  The red lines correspond to the transitions on the secondary diagonal in Fig.~\ref{quanTrans}, the black ones to the transitions between states $3$ and $9$.}
\label{floquet}
\end{figure}

Similar to the classical case, cf. subsection~\ref{chAA}, important qualitative features of the transition probability $P^Q_t(E_f|E_i)$ may be understood from an analysis of the system at constant $\la$. We are then concerned with a quantum system with time-periodic Hamiltonian that is most conveniently analyzed within Floquet theory \cite{Hol16,Shi65,Zel66, Sam73,FaiManRap78}. For a quantum system with time-periodic Hamiltonian, $\hH(t) = \hH(t+T)$, where $T:=2\pi/\om$ in our case, the states $\left|\psi_n(t)\right>$ have the general form 

\begin{equation}\label{defFloquetstates}
\left|\psi_n(t)\right> = \left|\varphi_n(t)\right> e^{-i\epsilon_n t}.
\end{equation}

Here the \textit{Floquet functions} $\left|\varphi_n(t)\right> $ share the periodic time-dependence with the Hamiltionian,
\begin{equation}
\left|\varphi_n(t)\right> = \left|\varphi_n(t+T)\right>, 
\end{equation}
and the quantity $\epsilon_n$ is known as {\em quasienergy}. Similar to quasimomentum in {\em spatially} periodic quantum systems quasienergies are only defined within integer multiples of $\om$ such that all values  
\begin{equation}\label{epsmult}
   \epsilon_{n,m} := \epsilon_n+ m\,\omega,\quad  m \in \mathbb{Z}
\end{equation}
are equivalent to each other. The main virtue of the Floquet states~\eqref{defFloquetstates} is that a general solution of the Schr\"odinger equation may be written as their superposition with {\em time-independent} coefficients. 

As a rule, the determination of the Floquet functions and their corresponding quasienergies can only be done numerically, e.g., from the relation 
\begin{equation}\label{Floquet}
\hat{U}(t+T,t) \left|\psi_n(t)\right> = e^{-i\epsilon_n T} \left|\psi_n(t)\right>
\end{equation}
characterizing the time evolution for one period of the driving. To do so in the present context we work in the basis of energy eigenstates $\phi_n$ of the undriven Hamiltonian $H_0$, i.e., for a given value of $\la$ we propagate the first forty states $\phi_n, \, n=1, \dots, 40$, for one period $T$ with the Crank-Nicolson-method, determine the matrix elements 
\begin{equation}
  U_{n,m} := \left<\phi_n\right|\hat{U}(T,0)\left|\phi_m\right>,
\end{equation}
and find the eigenvectors and eigenvalues~\eqref{Floquet} that fix the corresponding $\left|\varphi_n(t)\right>$ and $\epsilon_n$ according to~\eqref{defFloquetstates}. 

Fig.~\ref{floquet} shows results obtained in this way for the parameter values of Figs.~\ref{quanTrans} and \ref{quanWork}. All quasienergies have been mapped to the interval $0\leq \eps_n / \om< 1$ by appropriately chosen values of $m$ in~\eqref{epsmult}. The upper and the lower boundaries of the figure  corresponding to $\eps_n/\om=1$ and $\eps_n/\om=0$, respectively, have hence to be identified. Points arising from neighbouring values of $\la$ are connected if the absolute value of the scalar product between their corresponding Floquet functions is larger than with any other Floquet function. Note that {\em all} energy levels $E_n$ of the unperturbed system give rise to a quasienergy in the interval $[0,1)$, cf. \eqref{epsmult}. Therefore only a selection of lines is shown.  

For a slowly evolving amplitude $\la(t)$ the system will adiabatically follow the lines $\eps_n(\la)$ up to $\la=\lam$ and then go back again. At the various crossings or avoided crossings on its way it may perform transitions to other quasienergies and then end up in a different state when $\la=0$ at $t=\tf$. Whether such transitions occur or not depends on the specifics of both the crossing and the participating states and has to be carefully checked for each situation individually. 

But even without pinpointing the details at each crossing we may understand the transitions found or missing in Fig.~\ref{quanTrans} on the basis of  Fig~\ref{floquet} in a qualitative way. Let us focus first on the transition on the secondary diagonal. Their corresponding $\eps_n(\la)$-lines are shown in red in Fig.~\ref{floquet}. As can be seen they always come in pairs starting at $\la=0$ with the same quasienergy. This is simply a consequence of their initial {\em energies} differing by a multiple of $\om$. With increasing $\la$ they evolve along different lines and pick up different phases. Merging finally again for $t=\tf$ at the same value of $\eps_n$ these phases may induce constructive or destructive interference in this way deciding whether a transition occurs or not. 

This is a simple and robust transition mechanism since no level crossing is involved. All that is needed are two states of the undriven system with energies separated by a multiple of $\om$. It is rather reminiscent of the classical transitions discussed in subsection~\ref{chAA}. Nevertheless, a transition is not guaranteed. Depending on the detailed behaviour of the participating $\eps_n(\la)$-lines destructive interference a the final value $\la=0$ may suppress a transition that otherwise seems completely plausible. This is the reason for the absence of transitions between $i=16$ and $f=22$ and vice versa in Fig.~\ref{quanTrans}.

The transitions between $i=3$ and $f=9$ shown by the black lines in Fig.~\ref{floquet} are of different nature. This is already evident from the fact that the participating quasienergies do not coincide at $\la=0$. Correspondingly, $E_9-E_3$ is no multiple of $\om$. Let us assume that we start in $\left|\phi_{3}\right>$ at $t=0$. The state then closely follows the one emerging from this initial condition with hardly any additional component up to almost $\la=\lam$. However, near $\la\simeq 3.4$ there is an avoided crossing of quasienergies. In fact, the explicit calculation shows that the instantaneous {\em energies} of the states originating from $\left|\phi_{3}\right>$ and $\left|\phi_{9}\right>$ differ at $\la\simeq 3.4$ by just $4\om$. The initial wave packet splits in a generalized Landau-Zener transition and the new contribution to the state from $\left|\phi_{9}\right>$ remains present in the superposition all the way down back to $\la=0$. There it gives rise to a non-zero probability for $E_9$ in the second energy measurement. 

Let us at this point emphasize again that an identification of really occurring transitions solely on the basis of Fig.~\ref{floquet} is impossible. For each line $\eps_n(\la)$ there are rather many avoided crossings. Whether or not they really give rise to a transition with appreciable probability depends on the details of the associated Floquet state and the system state at the corresponding time, i.e. on information that goes well beyond to what is contained in Fig.~\ref{floquet}.

\section{The dependence on $\lam$}\label{sec:max}

\begin{figure}
\centering
\includegraphics[width=\linewidth]{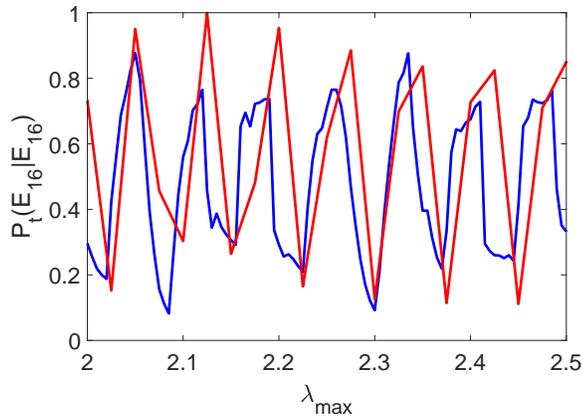}
\caption{Classical (blue) and quantum (red) transition probability $P(E_{16}|E_{16})$ as function of $\lam$. Parameter values are the same as in the other figures.}
\label{lambda}
\end{figure}

A peculiar feature of our system -- both classical and quantum mechanical -- is an oscillatory variation of the transition probabilities $P^C(E_f|E_i)$ and $P^Q(E_f|E_i)$, respectively, with the maximal amplitude $\lam$ of the driving. In Fig.~\ref{lambda} this is shown exemplarily for the case $i=f=16$ in the interval $2\leq \lam\leq 2.5$. The oscillations are clearly visible and classical and quantum results are in close correspondence. 

\begin{figure*}
\centering
\includegraphics[width=.4\linewidth]{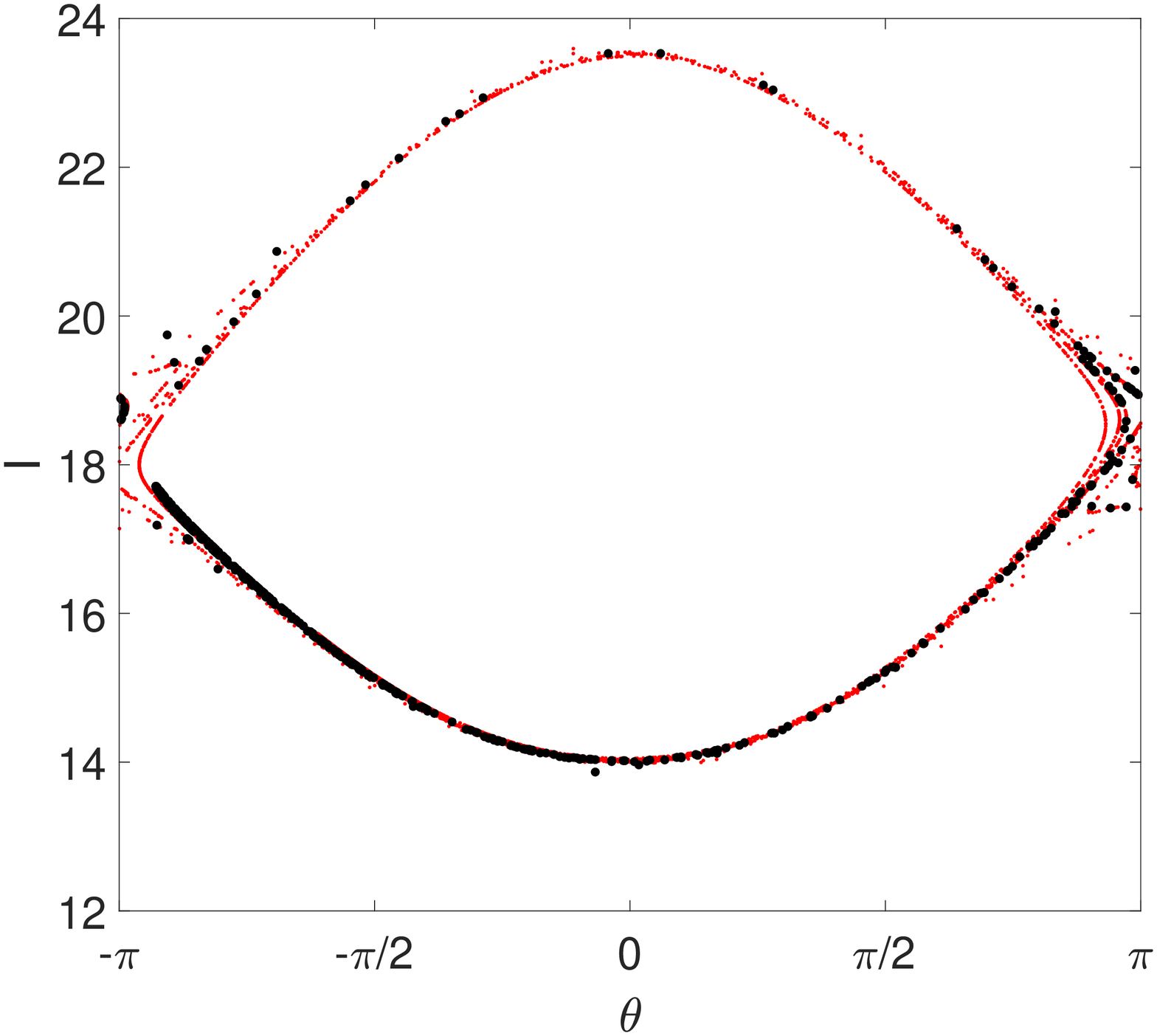}\hspace{.1\linewidth}
\includegraphics[width=.4\linewidth]{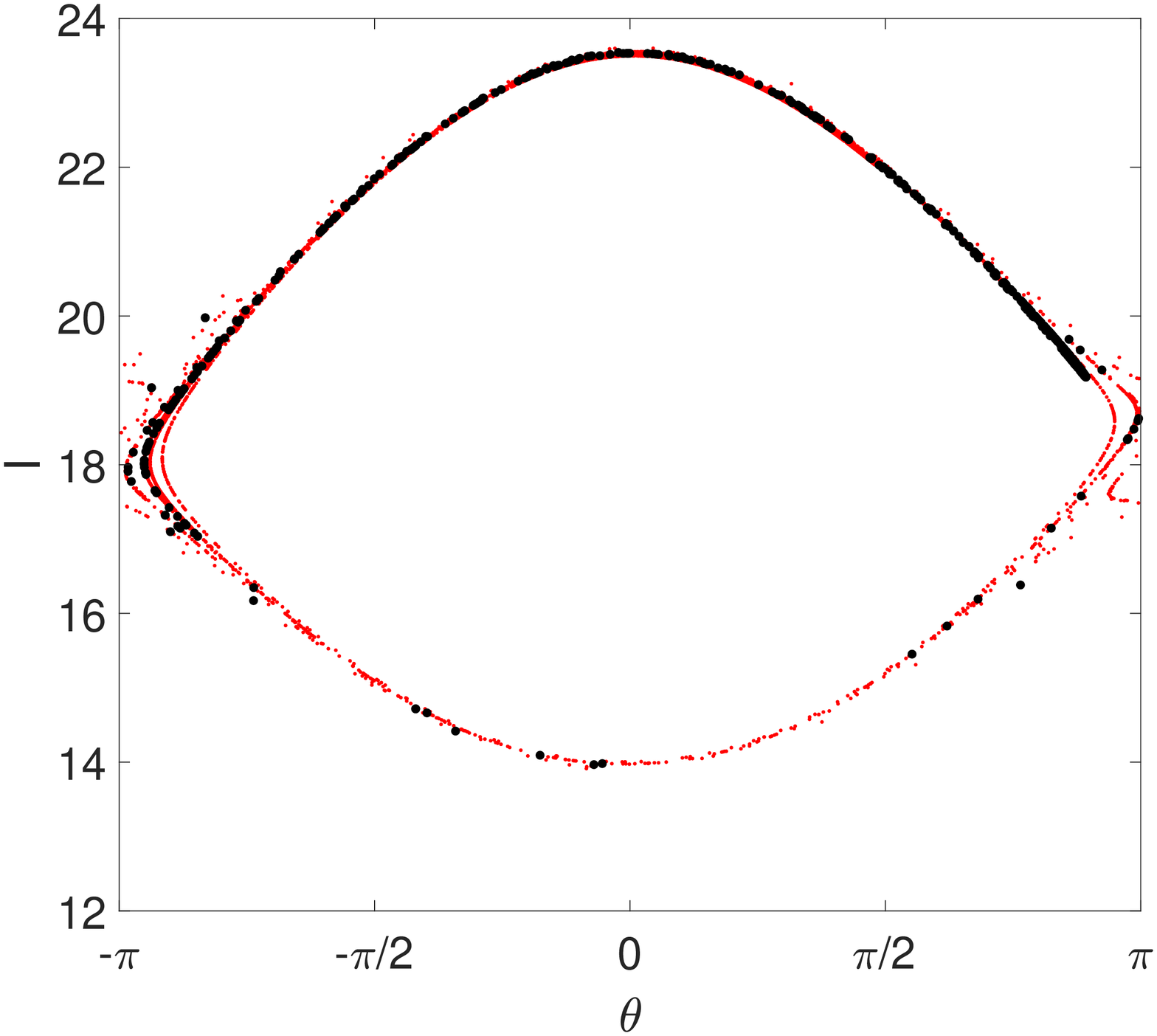}
\caption{Poincar\'e sections of classical orbits originating from $E_i=E_{16}$ (lower parts) and $E_i=E_{22}$ (upper parts) just before the two parts separate again in the course of decreasing $\la(t)$ for $\lam = 2.05$ (left) and $\lam = 2.085$ (right). For both initial energy values $2500$ classical trajectories were generated (red points), $500$ randomly selected initial conditions with $E_i=E_{16}$ give rise to the black dots. In the left figure almost all of these black points return to their initial energy value, in the right one most end up at $E_f=E_{22}$.}
\label{aufteilung}
\end{figure*}

For the quantum case there is simple explanation for this behaviour. As discussed already in section \ref{sec:Floquet} the quantum transition probability depends on whether the two participating Floquet states interfere constructively or destructively at the end of the driving process. Since these states collect different phases during the driving that depend on $\lam$ the oscillatory dependence seems natural. 

With interference of probability amplitudes being a genuine quantum phenomenon it is not obvious to find the mechanism behind these oscillations for the classical case. Nevertheless, it is rather analogous. 

In Fig.~\ref{aufteilung} we compare two Poincar\'e plots for the full dynamics with time-dependent $\la(t)$ with slightly different values of $\lam$. The upper and lower parts of the orbits derive from the initial energies $E_i=E_{16}$ and $E_i=E_{22}$, respectively. Shown is the situation exactly at the moment when the two parts detach at decreasing $\la$. The black dots correspond to a selection of systems that started with $E_i=E_{16}$ at $t=0$. 

For $\lam = 2.05$ (left part of the Figure) practically all these black dots are on the lower part of the curve and will therefore return to their initial energy value $E_f=E_{16}$. This corresponds to the first maximum of the blue line in Fig.~\ref{lambda}. If $\lam$ is only slightly larger, $\lam=2.085$, the dynamics are such that almost all black points are located on the upper part of the orbit just before the separation takes place. This is shown in the right part of Fig.~\ref{aufteilung}. These points, although started with $E_i=E_{16}$, will end up in a state with final energy $E_f=E_{22}$. Therefore, the transition probability $P^C(E_{16}|E_{16})$ will be small in accordance with the second minimum of the blue line in Fig.~\ref{lambda}. 

The different phases of the Floquet functions hence find their classical equivalent in the different number of circulations the phase space points undergo on the joint orbit at sufficiently large $\la$.

\section{Conclusion} \label{sec:conc}

The present paper was concerned with the classical and quantum mechanical analysis of a periodically driven anharmonic oscillator where the slowly varying amplitude of the driving smoothly increased from zero up to a maximum value and then returned back to zero in the end. The system is sufficiently simple to allow a rather detailed study both within classical and quantum mechanics. Nevertheless, it is representative for a whole class of nonlinear oscillators driven by time-periodic signals that are frequently subject of theoretical and experimental investigations. Our focus was on the possibilities to store or retrieve energy from the system in the course of driving, a question linked to the work statistics observed. This is of particular importance since the appropriate definition of work in small quantum systems is still controversial. Our analysis builds on numerical solutions of Hamiltons and Schr\"odingers equation of motion, respectively, complemented by approximate analytical results that establish ways to an intuitive understanding of the results. 
 
Decoupling the system from its surroundings during the driving the central quantity of interest is the transition probability $P(E_f|E_i)$ to end in a state with energy $E_f$ when started in one with energy $E_i$. Both classically and quantum mechanically this transition probability has a rather peculiar form. Only within a definite energy window around the resonance energy corresponding to the external periodic signal transition may occur with appreciable probability. For most initial energies within this window only transitions to one
particular final energy occur. 

Classically, this can be related to the mechanism of separatrix crossing that is most conveniently analyzed by transforming to action-angle variables of the undriven system. Employing the so-called pendulum approximation the results for the classical transition probability obtained from the numerical solution of the equations of motion can be reproduced rather well, in particular for small maximum amplitude of the driving. 

In the quantum setting most transitions can be traced back to constructive or destructive interference of Floquet states which the system follows adiabatically when the driving amplitude changes sufficiently slowly. Additional possibilities for transitions arise due to avoided crossing of quasienergy levels with associated generalized Landau-Zener transitions. 

We found a surprisingly close analogy between classical and quantum results. The overall shape of the transition probabilities is very similar and a pronounced oscillatory variation of the transition probability with the maximum value $\lam$ of the driving amplitude is found in both cases. For the quantum system this is due to constructive or destructive interference of Floquet states. The corresponding classical mechanism is related to integer or half-integer numbers of circulation of the phase space points on orbits originating from different energy values. 

There are, of course, also important differences. The energy window for off-diagonal transitions is smaller in the quantum case. This is mostly due to the discrete energy spectrum in the quantum case. Moreover, transitions that are impossible classically may occur for the quantum system due to avoided crossings of quasienergy levels at sufficiently large values of $\la$. 

Several interesting questions remain for further research. It would be very interesting to underpin the close correspondence between classical and quantum results by a semi-classical analysis. Also, quantization of the pendulum approximation may contribute to a quantitative understanding of the classical analoge of Floquet interferences.\\~\\

\begin{acknowledgments}
We would like to thank Martin Holthaus, Onno Diermann and Sebastian Rosmej and the members of the DFG Research Unit FOR2692 for fruitful discussions.
This work has been funded by the Deutsche Forschungsgemeinschaft (DFG, German Research Foundation) -- 397082825. 
\end{acknowledgments}



%

\end{document}